\documentclass{article}
\usepackage{amsfonts}
\usepackage{amsmath}
\usepackage{amssymb}
\usepackage{graphicx}
\providecommand{\U}[1]{\protect\rule{.1in}{.1in}}

\begin{document}

\title{ The Thermodynamic Covariance Principle}
\author{Giorgio SONNINO\\Department of Theoretical Physics and Mathematics\\ Universit\'{e} Libre de Bruxelles (ULB), Campus Plain CP 231 \\Boulevard de Triomphe, 1050 Brussels, Belgium.\\ Royal Military School (RMS),\\Av. de la Renaissance 30, 1000 Brussels, Belgium\\ Email: gsonnino@ulb.ac.be
\and Alberto SONNINO\\ Ecole Polytechnique de Louvain (EPL) \\ Universit{\'e} Catholique de Louvain (UCL), Rue Archim$\grave{\rm e}$de\\1 bte L6.11.01, 1348 Louvain-la-Neuve - Belgium. \\ Email: alberto.sonnino@gmail.com}
\maketitle

\begin{abstract}
The concept of {\it equivalent systems} from the thermodynamic point of view, originally introduced by Th. De Donder and I. Prigogine, is deeply investigated and revised. From our point of view, two systems are thermodynamically equivalent if, under transformation of the thermodynamic forces, both the entropy production and the Glansdorff-Prigogine dissipative quantity remain unaltered. This kind of transformations may be referred to as the {\it Thermodynamic Coordinate Transformations} (TCT). The general class of transformations satisfying the TCT is determined. We shall see that, also in the nonlinear region ({\it i.e.}, out of the Onsager region), the TCT preserve the reciprocity relations of the transformed transport matrix. The equivalent character of two transformations under TCT, leads to the concept of {\it Thermodynamic Covariance Principle} (TCP) stating that all thermodynamic equations involving the thermodynamic forces and flows ({\it e.g.}, the closure flux-force relations) should be covariant under TCT.

\end{abstract}

\section{Introduction}
Onsager's theory is based on three assumptions \cite{onsager1}-\cite{onsager2}: {\bf i)} {\it The probability distribution function for the fluctuations of thermodynamic quantities} (Temperature, pressure, degree of advancement of a chemical reaction etc.) {\it is a Maxwellian} {\bf ii)} {\it Fluctuations decay according to a linear law} and {\bf iii)} {\it The principle of the detailed balance} (or the microscopic reversibility) {\it is satisfied}.  Onsager showed the equivalence of the assumptions {\bf i)}-{\bf iii)} with the equations 
\begin{equation}\label{i1}
J_\mu=\tau_{0\mu\nu}X^\nu\qquad ;\qquad \frac{\partial\tau_{0\mu\nu}}{\partial X^\lambda}=0
\end{equation}
\noindent where $\tau_{0\mu\nu}$ are the transport coefficients and $X^\mu$ and $J_\mu$ denote the thermodynamic forces and the conjugate fluxes, respectively. Assumption {\bf iii)} allows deriving the {\it reciprocity relations} $\tau_{0\mu\nu}=\tau_{0\nu\mu}$. Note that in Eq.~(\ref{i1}), as well as in the sequel, the summation convention on the repeated indexes is understood. The Onsager theory of fluctuations starts from the Einstein formula linking the probability of a fluctuation, $\mathcal W$, with the entropy change, $\Delta S$, associated with the fluctuations from the state of equilibrium
\begin{equation}\label{i2a}
\mathcal{W}=W_0\exp[\Delta S/k_B]
\end{equation}
\noindent In Eq.~(\ref{i2a}), $k_B$ is the Bolzmann constant and $W_0$ is a normalization constant, which ensures that the sum of all probabilities equals to one. Prigogine generalized Eq.~(\ref{i2a}), which applies only to adiabatic or isothermal transformations, by introducing the entropy production due to fluctuations. Denoting by $\xi_i$ ($i=1\cdots m$) the $m$ deviations of the thermodynamic quantities from their equilibrium value, Prigogine proposed that the probability distribution of finding a state in which the values $\xi_i$ lie between $\xi_i$ and $\xi_i+d\xi_i$ is given by \cite{prigogine}
\begin{equation}\label{i3a}
\mathcal{W}=W_0\exp[\Delta_{\rm I}  S/k_B]\qquad\quad
{\rm where}\qquad \Delta_{\rm I}   S=\int_E^F d_{\rm I} s\quad  {\rm ;}\quad \frac{d_{\rm I}  s}{dt}\equiv\int_\Omega\sigma dv
\end{equation}
\noindent $dv$ is a (spatial) volume element of the system, and the integration is over the entire space $\Omega$ occupied by the system in question. $E$ and $F$ indicate the equilibrium state and the state to which a fluctuation has driven the system, respectively. Note that this probability distribution remains unaltered for flux-force transformations leaving invariant the entropy production. Concrete examples of chemical reactions, equivalent from the thermodynamic point of view, have also been analyzed in literature. As an example, among these, we choose the simplest of all. Let us consider, for example, a chemical system in which 

\noindent {\bf a)} two {\it isomerisations} $A\rightarrow B$ and $B\rightarrow  C$ take place \cite{prigogine}. 

\noindent From the macroscopic point of view, the chemical changes could be equally well described by the 

\noindent {\bf b)} two {\it isomerisations} $A\rightarrow  C$ and $B\rightarrow  C$. 

\noindent It can be checked that, under a linear transformation of the thermodynamic forces ({\it i.e.}, in this case, a linear transformation of the chemical affinities) the entropy productions, corresponding to the two chemical reactions {\bf a)} and {\bf b)}, are equal. Indeed, the corresponding affinities of the reactions {\bf a)} read : $A^1=\mu_A-\mu_B$, and $A^2=\mu_B-\mu_C$, with $A^i$ and $\mu_i$ ($i=A,B,C$) denoting the affinities and the chemical potentials, respectively. The change per unit time of the mole numbers is given by 
\begin{equation}\label{tcp1}
\frac{dn_A}{dt}=-v_1\ ;\qquad \frac{dn_B}{dt}=v_1-v_2\ ; \qquad \frac{dn_C}{dt}=v_2
\end{equation}
\noindent with $v_i$ ($i=1,2$) denoting the chemical reaction rates. In this case the thermodynamic forces and the flows are the chemical affinities (over temperature) and the chemical reaction rates, respectively {\it i.e.,} $X^\mu=A^\mu/T$ and $J_\mu=v_\mu$. Hence, the corresponding entropy production reads $d_{I} S/dt=A^1/Tv_1+A^2/Tv_2>0$. The affinities corresponding to reactions {\bf b)} are related to the old ones by 
\begin{equation}\label{tcp2}
A'^1=\mu_A-\mu_C=A^1+A^2\ ; \qquad A'^2=\mu_B-\mu_C=A^2
\end{equation}
\noindent By taking into account that 
\begin{equation}\label{tcp3}
\frac{dn_A}{dt}=-v'_1\ ;\qquad \frac{dn_B}{dt}=-v'_2\ ; \qquad \frac{dn_C}{dt}=v'_1+v'_2
\end{equation}
\noindent we get
\begin{equation}\label{tcp4}
v_1=v'_1\ ;\qquad v_2=v'_1+v'_2
\end{equation}
\noindent where the invariance of the entropy production is manifestly shown. Indeed,
\begin{equation}\label{tcp5}
\!\!\!\! d_{I} S/dt=(A^1/T)v_1+(A^2/T)v_2=(A'^1/T)v'_1+(A'^2/T)v'_2=d_{I} S'/dt
\end{equation}
\noindent or $J_\mu X^\mu = J'_\mu X'^{\mu}$ (where the Einstein summation convention on the repeated indexes is adopted). On the basis of the above observations, Th. De Donder and I. Prigogine formulated, for the first time, the concept of {\it equivalent systems from the thermodynamical point of view}. For Th. De Donder and I. Prigogine, {\it thermodynamic systems are thermodynamically equivalent if, under transformation of fluxes and forces, the bilinear form of the entropy production remains unaltered, i.e.,} $\sigma=\sigma'$. \cite{prigogine2}.

\noindent However, the condition of the invariance of the entropy production is not sufficient to ensure the equivalence character of the  two descriptions $(J_\mu , X^\mu)$ and $(J'_\mu , X'^\mu )$. Indeed, we can convince ourselves that there exists a large class of transformations such that, even though they leave unaltered the expression of the entropy production, they may lead to certain paradoxes to which Verschaffelt has called attention \cite{verschaffelt}-\cite{davies}. This obstacle can be overcome if one takes into account one of the most fundamental and general theorems valid in thermodynamics of irreversible processes : the {\it Universal Criterion of Evolution}. In general, Glansdorff and Prigogine have shown that : {\it For time-independent boundary conditions, a thermodynamic system, even in strong non-equilibrium conditions, relaxes to a stable stationary state in such a way that the following Universal Criterion of Evolution is satisfied} \cite{glansdorff1}-\cite{glansdorff2} : 
\begin{equation}\label{tcp5a}
\int_\Omega J_\mu\frac{\partial X}{\partial t}^\mu\ dV\leq 0
\end{equation}
\noindent Here, $\Omega$ is the volume occupied by the system and $dV$ the volume-element, respectively. In addition
\begin{equation}\label{tcp5b}
 \int_\Omega J_\mu\frac{\partial X}{\partial t}^\mu dV= 0 \quad {\rm At\ the\ steady\ state} 
 \end{equation}
\noindent Quantity $P\equiv\int_\Omega J_\mu\frac{\partial X}{\partial t}^\mu dV$ may be referred to as {\it the Glansdorff-Prigogine dissipative quantity}.

\noindent Let us check the validity of this theorem by considering two, very simple, examples. Let us consider, for instance, a closed system containing $m$ components ($i=1\dots m$) among which chemical reactions are possible. The temperature, $T$, and the pressure, $p$, are supposed to be constant in time. The chance in the number of moles $n_i$, of component $i$, is 
\begin{equation}\label{tc1}
 \frac{dn_i}{dt}=\nu_i^jv_j
 \end{equation}
\noindent with $\nu_i^j$ denoting the stoichiometric coefficients. By multiplying both members of Eq.~(\ref{tc1}) by the time derivative of the chemical potential of component $i$, we get
\begin{equation}\label{tc2}
{\dot\mu}^i\frac{dn_i}{dt}=\Bigl(\frac{\partial\mu^i}{\partial n_\kappa}\Bigr)_{pT}\frac{dn_i}{dt}\frac{dn_\kappa}{dt}=\nu_i^j{\dot\mu}^iv_j
 \end{equation}
\noindent By taking into account the De Donder law between the affinities $A^i$ and the chemical potentials {\it i.e.}, $A^j=-\nu_i^j\mu^i$, we finally get
\begin{equation}\label{tc3}
P\equiv J_\mu\frac{dX^\mu}{dt}=v_j\frac{d}{dt}\Bigl(\frac{A^j}{T}\Bigr)
=-\frac{1}{T}\Bigl(\frac{\partial\mu^i}{\partial n_\kappa}\Bigr)_{pT}\frac{dn_i}{dt}\frac{dn_\kappa}{dt}\leq 0
\end{equation}
\noindent where the negative sign of the term on the right-hand side is due to the second law of thermodynamics. Hence, the Glansdorff-Prigogine dissipative quantity $P$ is always negative, and it vanishes at the stationary state.

\noindent As a second example, we analyze the case of heat conduction in non-expanding solid. In this case the thermodynamic forces and the conjugate flows are the (three) components of the gradient of the inverse of the temperature, $X^\mu=[{\rm grad}\ (1/T)]^\mu$ and the (three) components of the heat flow, $J_\mu={\bf J}_{q\mu}$ (with i=1,2,3), respectively. Hence, 
\begin{equation}\label{tcp6}
P\equiv\int_\Omega J_\mu \frac{\partial X^\mu}{\partial t}\ dV=\int_\Omega {\bf J}_{q\mu}\frac{\partial}{\partial t}[{\rm grad}\ (1/T)]^\mu\ dV
\end{equation}
\noindent With partial integration, and by taking into account the energy law  
\begin{equation}\label{tcp7}
\rho c_v\frac{\partial T}{\partial t}=-{\rm div}\ {\bf J}_q 
\end{equation}
\noindent after simple calculations, we get
\begin{equation}\label{tcp8}
P=-\int_\Omega\frac{\rho c_v}{T^2}\Bigl(\frac{\partial T}{\partial t}\Bigr)^2 dV\leq 0
\end{equation}
\noindent with $P=0$ at the steady state. In Eqs~(\ref{tcp7}) and (\ref{tcp8}), $\rho$ and $c_v$ are the mass density and the specific heat at volume constant, respectively. By summarizing, without using neither the Onsager reciprocal relations nor the assumption that the phenomenological coefficients (or linear phenomenological laws) are constant, the dissipative quantity $P$ is always a negative quantity. This quantity vanishes at the stationary state. In the two above-mentioned examples, the thermodynamic forces are the chemical affinities (over temperature) and the gradient of the inverse of temperature, respectively. However, we could have adopted a different choice of the thermodynamic forces. If we analyze, for instance, the case of heat conduction in non-expanding solid, where chemical reactions take place simultaneously, we can choose as thermodynamic forces a combination of the (dimensionless) chemical affinities (over temperature) and the (dimensionless) gradient of the inverse of temperature. Clearly, this representation is  thermodynamically equivalent to the one where the thermodynamic forces are simply the chemical affinities (over temperature) and the gradient of the inverse of temperature, only if the transformation between the thermodynamic forces preserves the negative sign of the quantity $P$. In particular, the two equations for the stationary states [the Eq.~(\ref{tcp5b}) and its transformed] must admit the {\it same} solutions. 

\noindent To sum up, the (admissible) thermodynamic forces should satisfy the following two conditions:

\noindent {\bf 1)} {\it The entropy production, $\sigma$, should be invariant under transformation of the thermodynamic forces $\{X^\mu\}\rightarrow\{X^{\prime\mu}\}$} and

\noindent {\bf 2)} {\it The Glansdorff-Prigogine dissipative quantity, $P$, should also remain invariant under the forces transformations $\{X^\mu\}\rightarrow\{X^{\prime\mu}\}$}.

\noindent Condition {\bf 2)} derives from

\noindent {\bf 2a)} {\it A {\bf steady state} should be transformed into a {\bf steady state}} and

\noindent {\bf 2b)} {\it A {\bf stable} steady state should be transformed into a {\bf stable} state state, with the same "degree" of stability}.

\noindent Magnetically confined tokamak plasmas are a typical example of thermodynamic systems, out of Onsager's region, where the Glansdorff-Prigogine dissipative quantity remains invariant under transformation of the thermodynamic forces \cite{sp}. Conditions {\bf 1)} and {\bf 2)} allow determining univocally the class of admissible thermodynamic forces \cite{sonnino}. This kind of transformations may be referred to as the {\it Thermodynamic Coordinate Transformations} (TCT).

\noindent In the light of the above, we formulate the following principle : {\it Two systems are equivalent from the thermodynamic point of view if, under transformation $\{X^\mu\}\rightarrow \{X^{\prime\mu}\}$, we have $\sigma=\sigma'$ and $P=P'$} \cite{sonnino}. The aim of this work is to derive from the {\it admissible} thermodynamic forces, the most general class of transformations (TCT) so that two thermodynamic systems are equivalent from the thermodynamic point of view. We shall prove that under TCT, $\sigma$ and $P$ remain invariant and, in addition, the reciprocity relations of the transformed transport matrix are preserved.

\noindent The thermodynamic equivalence principle leads, naturally, to the following {\it Thermodynamic Covariance Principle} (TCP) : {\it The nonlinear closure equations ({\it i.e.} the flux-force relations valid out of Onsager's region) must be covariant under} (TCT). The essence of the TCP is the following. The equivalent character of two representations is warranted if, and only if, the fundamental thermodynamic equations are covariant under TCT. In fact, a covariant formulation guarantees the writing of the fundamental laws of thermodynamics (for instance, the flux-force closure equations) in a form that is manifestly invariant under transformation between the admissible thermodynamic forces. This is the correct mathematical formalism to ensure the equivalence between two different representations. Note that the TCP is trivially satisfied by the closure equations valid in the Onsager region. It is worthwhile mentioning that the linear version of the TCT (see the next section) is actually largely used in a wide variety of thermodynamic processes ranging from non equilibrium chemical reactions to transport processes in tokamak plasmas (see, for examples, the papers cited in the book \cite{balescu2}). To the authors knowledge, the validity of the thermodynamic covariance principle has been verified empirically without exception in physics until now.

\section{The Thermodynamic Covariance Principle}\label{isoef}
\vskip 0.2truecm

\noindent Consider a thermodynamic system driven out of equilibrium by a set of $n$ independent thermodynamic forces $\{X^\mu\}$ ($\mu=1,\cdots n$). It is also assumed that the system is submitted to time-independent boundary conditions. The set of conjugate flows, $\{J_{\mu}\}$, is coupled to the thermodynamic forces through the relation 
\begin{equation}\label{ief}
J_{\mu}=\tau_{\mu\nu}(X)X^\nu
\end{equation}
\noindent where the transport coefficients, $\tau_{\mu\nu}(X)$, may depend on the thermodynamic forces. The symmetric piece of $\tau_{\mu\nu}(X)$ is denoted with $g_{\mu\nu}(X)$ and the skew-symmetric piece as $f_{\mu\nu}(X)$:
\begin{equation}\label{ief1}
\tau_{\mu\nu}(X)=\frac{1}{2}[\tau_{\mu\nu}(X)+\tau_{\nu\mu}(X)]+\frac{1}{2}[\tau_{\mu\nu}(X)-\tau_{\nu\mu}(X)]=g_{\mu\nu}(X)+f_{\mu\nu}(X)
\end{equation}
\noindent where
\begin{eqnarray}\label{ief2}
&&g_{\mu\nu}(X)\equiv\frac{1}{2}[\tau_{\mu\nu}(X)+\tau_{\nu\mu}(X)]=g_{\nu\mu}(X)\\
&&f_{\mu\nu}(X)\equiv\frac{1}{2}[\tau_{\mu\nu}(X)-\tau_{\nu\mu}(X)]=-f_{\nu\mu}(X)
\end{eqnarray}
\noindent  It is assumed that $g_{\mu\nu}(X)$ is a positive definite matrix. For the sake of conciseness, in the sequel we drop the symbol $(X)$ in $\tau_{\mu\nu}$, $g_{\mu\nu}$ and $f_{\mu\nu}$, being implicitly understood that these matrices may depend on the thermodynamic forces. With the elements of the transport coefficients two objects are constructed: {\it operators}, which may act on thermodynamic tensorial objects and {\it thermodynamic tensorial objects}, which under coordinate (forces) transformations, obey to well specified transformation rules.
\vskip 0.2truecm
\noindent{\bf Operators}
\vskip 0.2truecm
Two operators are introduced, the {\it entropy production operator} $\sigma (X)$ and the {\it dissipative quantity operator} ${\tilde P}(X)$, acting on the thermodynamic forces in the following manner
\begin{eqnarray}\label{o1}
&&\sigma (X):\rightarrow\sigma(X)\equiv X gX^T\nonumber\\
&& {\tilde P}(X):\rightarrow {\tilde P}(X)\equiv X\tau \Bigl[\frac{dX}{d\varsigma}\Bigr]^T
\end{eqnarray}
\noindent In Eqs~(\ref{o1}), the transport coefficients are then considered as elements of the two $n$ x $n$ matrices, $\tau$ and $g$. Symbol $T$ stands for the transpose of the matrix. The positive definiteness of the matrix $g_{\mu\nu}$ ensures the validity of the second principle of thermodynamics: $\sigma\ge0$. These matrices multiply the thermodynamic forces $X$ expressed as $n$ x $1$ column matrices. We already anticipate that parameter $\varsigma$, defined in Eq.~(\ref{tct4}), is invariant under the thermodynamic coordinate transformations. Thermodynamic states $X_{s}$ such that
\begin{eqnarray}\label{o2}
\Big[{\tilde P}(X)\frac{d\varsigma}{dt}\Big]_{X=X_{s}}
\!\!\!\!=0
\end{eqnarray}
\noindent are referred to as {\it steady states}. Of course, the steady states should be invariant expressions under the thermodynamic coordinate transformations. Eqs~(\ref{o1}) {\it should not} be interpreted as the metric tensor $g_{\mu\nu}$, which acts on the coordinates. The metric tensor {\it acts only on} elements of the tangent space (like $dX^\mu$, see the forthcoming sub-section) or on the thermodynamic tensorial objects.
\vskip 0.2truecm
\noindent{\bf Transformation Rules of Entropy Production, Forces, and Flows}
\vskip 0.2truecm
In the previous section we stated that two systems are equivalent from the thermodynamic point of view if, under TCT, $\sigma=\sigma'$ {\it and} ${\tilde P}={\tilde P}'$ remain unaltered. In mathematical terms, this implies:
 \begin{equation}\label{tr1}
 \sigma=J_\mu X^\mu=J'_\mu X'^\mu
 \end{equation}
 \noindent This condition and the condition that also the dissipative quantity [cf. Eqs~(\ref{o1})] must be an invariant expression, require that the transformed thermodynamic forces and flows satisfy the relation  
\begin{eqnarray}\label{tr2}
&&X'^\mu=\frac{\partial X'^{\mu}}{\partial X^\nu} X^\nu\nonumber\\
&& J'_\mu=\frac{\partial X^{\nu}}{\partial X'^\mu}J_\nu
\end{eqnarray}
\noindent These transformations may be referred to as {\it Thermodynamic Coordinate Transformations} (TCT). The expression of entropy production becomes accordingly
 \begin{equation}\label{tr3}
 \sigma=J_\mu X^\mu=\tau_{\mu\nu}X^\mu X^\nu=g_{\mu\nu}X^\mu X^\nu=g'_{\mu\nu}X'^\mu X'^\nu=\sigma'
 \end{equation}
 \noindent From Eqs~(\ref{tr2}) and (\ref{tr3}) we derive
 \begin{equation}\label{tr4}
g'_{\lambda\kappa}=g_{\mu\nu}\frac{\partial X^\mu}{\partial X'^\lambda}
 \frac{\partial X^\nu}{\partial X'^\kappa}
\end{equation}
\noindent Moreover, inserting Eqs~(\ref{tr2}) and Eq.~(\ref{tr4}) into relation $J_{\mu}=(g_{\mu\nu}+f_{\mu\nu})X^\nu$, we obtain
\begin{equation}\label{tr5}
J'_\lambda=\Bigl(g'_{\lambda\kappa}+f_{\mu\nu}\frac{\partial X^\mu}{\partial X'^\lambda}\frac{\partial X^\nu}{\partial X'^\kappa}\Bigr)X'^\kappa
\end{equation}
\noindent or 
 \begin{equation}\label{tr6}
J'_\lambda=(g'_{\lambda\kappa}+f'_{\lambda\kappa})X'^\kappa\quad\qquad{\rm with}\qquad
f'_{\lambda\kappa}=f_{\mu\nu}\frac{\partial X^\mu}{\partial X'^\lambda}
 \frac{\partial X^\nu}{\partial X'^\kappa}
\end{equation}
\noindent Hence, the transport coefficients transform like a {\it thermodynamic tensor of second order}
\footnote{We may qualify as {\it thermodynamic tensor} or, simply {\it thermo-tensor}, (taken as a single noun) a set of quantities where {\it only transformations Eqs~(\ref{tr2}) are involved}. This is in order to qualify as a {\it tensor}, a set of quantities, which satisfies certain laws of transformation when the coordinates undergo a general transformation. Consequently every tensor is a thermodynamic tensor but the converse is not true.\label{thermotensor}}. It is easily checked that transformations (\ref{tr2}) preserve the validity of the reciprocal relations for transport coefficients {\it i.e.}, if $g_{\mu\nu}=g_{\nu\mu}$ then $g'_{\mu\nu}=g'_{\nu\mu}$ (and, if $f_{\mu\nu}=-f_{\nu\mu}$ then $f'_{\mu\nu}=-f'_{\nu\mu}$).

\vskip0.5truecm
\noindent{\bf Properties of the TCT}
\vskip0.5truecm

\noindent By direct inspection, it is easy to verify that the general solutions of equations (\ref{tr2}) are 
\begin{equation}\label{tct1}
X'^\mu=X^1F^\mu\Bigl(\frac{X^2}{X^1},\ \frac{X^3}{X^2},\ \cdots\ \frac{X^n}{X^{n-1}}\Bigr)
\end{equation}
\noindent where $F^\mu$ are {\it arbitrary functions} of variables $X^j/X^{j-1}$ with ($j=2,\dots, n$). Eq.~(\ref{tct1}) is the most general class of transformations expressing the equivalent character of two descriptions $\{X^\mu\}$ and $\{X^{\prime\mu}\}$. Hence, the TCT may be {\it highly nonlinear coordinate transformations} but, in the Onsager region, we may (or we must) require that they have to reduce to 
\begin{equation}\label{tct1a}
X'^\mu=c_\nu^\mu X^\nu
\end{equation}
\noindent where $c_\nu^\mu$ are constant coefficients (i.e., independent of the thermodynamic forces). The linear and homogeneous transformations (\ref{tct1a}) are largely used in literature because, besides their simplicity, they are {\it nonsingular} transformations. We note that from Eq.(\ref{tr2}), the following important identities are derived
\begin{equation}\label{tct2}
X^\nu\frac{\partial^2X'^\mu}{\partial X^\nu\partial X^\kappa}=0\qquad ;\qquad X'^\nu\frac{\partial^2X^\mu}{\partial X'^\nu\partial X'^\kappa}=0
\end{equation}
\noindent Moreover
\begin{eqnarray}\label{tct3}
&&dX'^\mu=\frac{\partial X'^{\mu}}{\partial X^\nu} dX^\nu \nonumber\\
&&\frac{\partial}{\partial X'^\mu}= \frac{\partial X^{\nu}}{\partial X'^\mu}\frac{\partial}{\partial X^\nu} 
\end{eqnarray}
\noindent i.e., $dX^\mu$ and $\partial/\partial X^{\mu}$ transform like a thermodynamic contra-variant and a thermodynamic covariant vector, respectively. According to Eq.~(\ref{tct3}), thermodynamic vectors $dX^\mu$ define the {\it tangent space} to $Ts$. It also follows that the operator $P(X)$, i.e. the dissipation quantity, and in particular the definition of steady states, are invariant under TCT. Parameter $\varsigma$, defined as 
\begin{equation}\label{tct4}
d\varsigma^2=g_{\mu\nu}dX^\mu dX^\nu
\end{equation}
\noindent is a scalar under TCT. The operator $\mathcal{O}$
\begin{equation}\label{pts14d}
\mathcal{O}\equiv X^\mu\frac{\partial}{\partial X^\mu}=X'^\mu\frac{\partial}{\partial X'^\mu}=\mathcal{O}'
\end{equation}
\noindent is also invariant under TCT. This operator plays an important role in the formalism \cite{sonnino}.

\noindent We may now enunciate the {\it Thermodynamic Covariance Principle} (TCP) : {\it All thermodynamic equations involving the thermodynamic forces and flows (e.g., the closure equations) should be covariant under} (TCT). 

\noindent From this principle, it is possible to obtain the (nonlinear) closure equations by truncating the equations (obtained, for instance, by kinetic theory) relating the thermodynamic forces with the conjugate flows in such a way that the resulting expressions satisfy the TCP. A concrete example is briefly mentioned in the Conclusions.

\bigskip

\section{ Conclusions}
If one requires only the entropy production to be invariant, but does not impose the auxiliary condition that he Glansdorff-Prigogine dissipative quantity should also be invariant under TCT, there exists the larger class of transformations leading to the phenomenological coefficients scheme, for which the reciprocal relations are not valid \cite{degroot}. As rightly pointed out by Verschaffelt and Davies, "to impose that only the entropy production must be invariant under flux-force transformations may lead to paradoxes or inconsistencies" \cite{verschaffelt}-\cite{davies}. We must also keep in mind that the original form of the entropy production, as derived from the various balance equations, is such that the Onsager relations are valid. The correct way to overcome this {\it impasse} is to require that both the entropy production and the Glansdorff-Prigogine dissipative quantity remain unaltered under TCT. 

\noindent The Nonlinear theory, the {\it Thermodynamic Field Theory} (TFT), reported in \cite{sonnino}, is based on the validity of the Thermodynamic Covariance Principle. This assumption allows truncating, at the lowest order, the (highly nonlinear) equations relating the thermodynamic forces with the conjugate flows. For example, in the case of magnetically confined tokamak-plasmas, these relations are obtained by kinetic theory and are expressed by highly nonlinear integral equations \cite{balescu1}. In this specific case, the generalized frictions are the thermodynamic forces whereas the conjugate flows are the Hermitian moments \cite{balescu2}. The Hermitian moments are linked to the deviation of the distribution function from the (local) Maxwellian, whereas the generalized frictions are linked to the collision term. According to the TCP, these integral equations may be truncated in such a way that the resulting expressions are covariant under TCT. The derived closure equations have been recently used for estimating the electron and ion losses (particle and heat losses) as well as the electron and ion distributions functions \cite{sp}-\cite{sonnino2}. The theoretical predictions are fairly in line with the preliminary experimental data obtained for FTU-plasmas (FTU= Frascati Tokamak Upgrade). 

\noindent We conclude by mentioning that attempts to derive a {\it generally covariant} thermodynamic field theory (GTFT) can be found in refs \cite{se1}-\cite{se2} and \cite{m}. In Refs~\cite{se1}-\cite{se2}, the general covariance has been assumed to be valid for general transformations in the space of thermodynamic configurations, whereas in \cite{m} it is argued that the entropy production rate should be invariant under general spatial coordinate transformations (in \cite{m} the invariance of the Prigogine-Glansdorff dissipation is not satisfied). However, it is a matter of fact that a {\it generally covariant} thermodynamic theory, which is based solely on the invariance of the entropy production is, in reality, respected only by a very limited class of thermodynamic processes. As mentioned, a correct thermodynamic formalism must satisfy the TCP.


\bigskip
\end{document}